\documentclass[twocolumn,showkeys,aps,prb,showpacs]{revtex4-1}
\UseRawInputEncoding
\usepackage{graphicx}
\usepackage[CJKbookmarks,dvipdfm,colorlinks,linkcolor=red,citecolor=blue]{hyperref}

\begin{document}

\title{Correlation-driven threefold  topological phase transition in monolayer $\mathrm{OsBr_2}$}

\author{San-Dong Guo$^{1}$, Wen-Qi Mu$^{1}$  and  Bang-Gui Liu$^{2,3}$}
\affiliation{$^1$School of Electronic Engineering, Xi'an University of Posts and Telecommunications, Xi'an 710121, China}
\affiliation{$^2$ Beijing National Laboratory for Condensed Matter Physics, Institute of Physics, Chinese Academy of Sciences, Beijing 100190, People's Republic of China}
\affiliation{$^3$School of Physical Sciences, University of Chinese Academy of Sciences, Beijing 100190, People's Republic of China}
\begin{abstract}
Spin-orbit coupling (SOC) combined with electronic correlation can induce topological phase transition, producing novel electronic states.
Here,  we investigate the impact of SOC combined with correlation effects
on  physical properties of monolayer $\mathrm{OsBr_2}$, based on  first-principles calculations with  generalized gradient
approximation plus $U$ (GGA+$U$) approach. With intrinsic out-of-plane magnetic anisotropy, $\mathrm{OsBr_2}$ undergoes threefold  topological phase transition with increasing $U$, and  valley-polarized quantum anomalous Hall insulator (VQAHI) to  half-valley-metal (HVM) to  ferrovalley insulator (FVI) to HVM to VQAHI to HVM to FVI transitions can be induced. These topological phase transitions are connected with
sign-reversible  Berry curvature  and  band inversion  between $d_{xy}$/$d_{x^2-y^2}$ and $d_{z^2}$ orbitals.
Due to $\bar{6}m2$ symmetry, piezoelectric polarization of  $\mathrm{OsBr_2}$ is  confined along the in-plane armchair direction, and only one $d_{11}$ is independent. For a given material, the correlation strength should be  fixed, and $\mathrm{OsBr_2}$ may be a piezoelectric VQAHI (PVQAHI), piezoelectric HVM (PHVM) or piezoelectric FVI (PFVI). The valley polarization
can be flipped  by reversing the magnetization of Os atoms, and the ferrovalley  (FV) and  nontrivial topological properties will be suppressed   by manipulating out-of-plane magnetization to in-plane one. In considered reasonable $U$ range, the estimated  Curie temperatures all are higher than room temperature.  Our findings provide a comprehensive understanding on possible electronic states of  $\mathrm{OsBr_2}$, and confirm that strong SOC  combined with electronic correlation can induce multiple quantum phase transition.

\end{abstract}
\keywords{Correlation, SOC, Phase transition, Piezoelectricity ~~~~~~~~~~~~~~~~~~~~~~~~~Email:sandongyuwang@163.com}

\maketitle

\section{Introduction}
The effect of SOC on  physical
properties of  materials has attracted intensive attention.  The SOC  plays a key role for magnetocrystalline anisotropy, topological and valley physics\cite{q1,q2,q2-1,q2-2,q3,q4}. For example,  two-dimensional (2D) materials with long-range ferromagnetic (FM) order can be
stabilized by SOC-induced magnetocrystalline anisotropy\cite{q1}. The  SOC can also stabilize the
topological phases by a nontrivial
bandgap in  quantum spin Hall insulator (QSHI) and quantum anomalous Hall insulator (QAHI)\cite{q2,q2-1,q2-2}.
The SOC can induce spontaneous valley polarization in  2D magnetic semiconductors with special crystal symmetry, like 2H-$\mathrm{VSe_2}$\cite{q4}.

The impact of electronic correlation on material properties has been a research hotspot\cite{q5,q6,q7,q8}. Correlation-driven topological phases have been achieved  in magic-angle twisted bilayer graphene\cite{q5}. Correlation-driven eightfold magnetic anisotropy can be realized in a 2D oxide monolayer\cite{q6}. Topological Fermi surface transition  can be induced by  varied electronic correlation in FeSe\cite{q7}. Coulomb interaction can induce quantum anomalous Hall (QAH) phase  in (111) bilayer of $\mathrm{LaCoO_3}$\cite{q8}.
Generally, the correlation effects are resultful in transition metal
elements with localized $d$ electrons. For special crystal symmetry, electron correlation can dramatically enhance SOC
effect of light elements in certain partially occupied orbital multiplets\cite{q9}.

\begin{figure}
  \includegraphics[width=7cm]{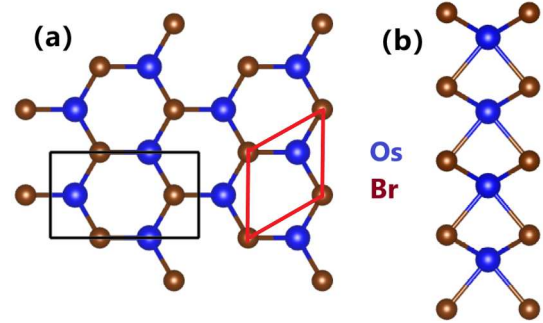}
  \caption{(Color online)The (a) top view and (b) side view of  crystal structure of $\mathrm{OsBr_2}$ monolayer, and the rhombus primitive cell (rectangle supercell) is marked by the red (black) frame. }\label{st}
\end{figure}
Recently,  correlation-induced novel electronic states, such as VQAHI and HVM,   have been predicted in many 2D materials\cite{q10,q11,q12,q13}.
For FeClF or $\mathrm{FeCl_2}$ monolayer, increasing electron correlation can induce FVI to HVM to VQAHI to HVM to FVI transitions with fixed out-of-plane magnetic anisotropy\cite{q10,q12}. However, with intrinsic magnetic anisotropy,  no special VQAHI and HVM states exist in FeClF monolayer, which means that  no topological phase transition is induced.
For monolayer $\mathrm{RuBr_2}$,  with fixed out-of-plane magnetic anisotropy, the phase diagram is the same with that of FeClF monolayer\cite{q13}.
The intrinsic phase diagram shows  VQAHI and HVM states, but only one HVM state can exist.
From FeClF to $\mathrm{RuBr_2}$, these differences are because Ru atom  has heavier atomic mass than Fe atom, which will lead to stronger SOC effects.
The Os atom has more stronger SOC effects than Ru atom, which may give rise to other novelty effects. Recently, monolayer $\mathrm{OsBr_2}$ with $1H$-$\mathrm{MoS_2}$ type structure is predicted to stable\cite{q14}.

In this work, the electronic correlation effects  on  electronic structures of $\mathrm{OsBr_2}$ monolayer are carefully investigated.
Different from  FeClF or $\mathrm{FeCl_2}$ or $\mathrm{RuBr_2}$, increasing electron correlation induces  threefold  topological phase transition in monolayer $\mathrm{OsBr_2}$  with intrinsic out-of-plane magnetic anisotropy, which means that there are three HVM states and  two VQAHI regions.
Due to missed centrosymmetry, $\mathrm{OsBr_2}$ is piezoelectric, and its piezoelectric properties are investigated.
The combination of piezoelectricity, topology and/or FV  in $\mathrm{OsBr_2}$ monolayer provides  a potential platform for multi-functional spintronic applications, and our works provide possibility to use the piezoelectric effect to control QAH or anomalous valley Hall effect.

\begin{figure}
  \includegraphics[width=8cm]{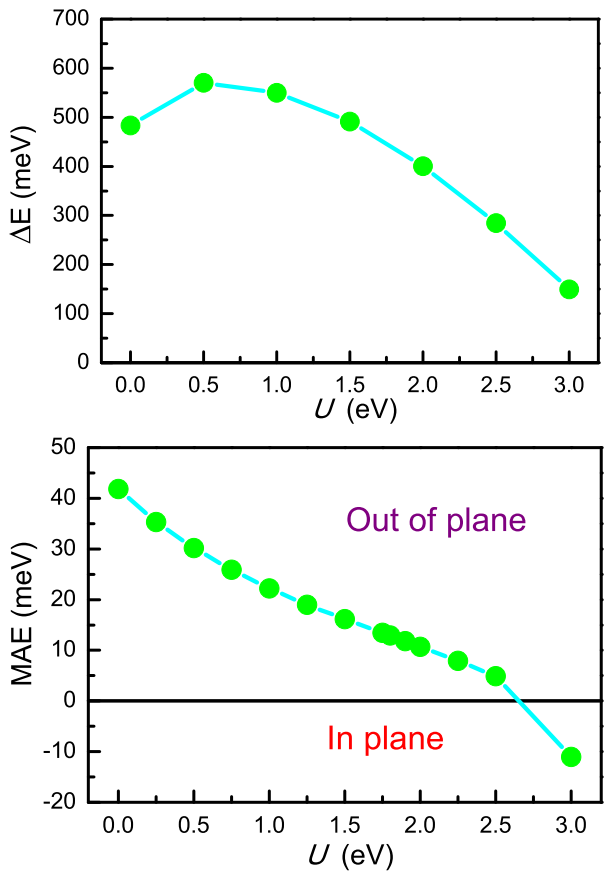}
  \caption{(Color online)The  energy differences $\Delta E$ between  AFM and FM ordering  and  MAE  of   $\mathrm{OsBr_2}$ monolayer as a function of $U$.}\label{u-em}
\end{figure}

\begin{figure*}
  \includegraphics[width=16cm]{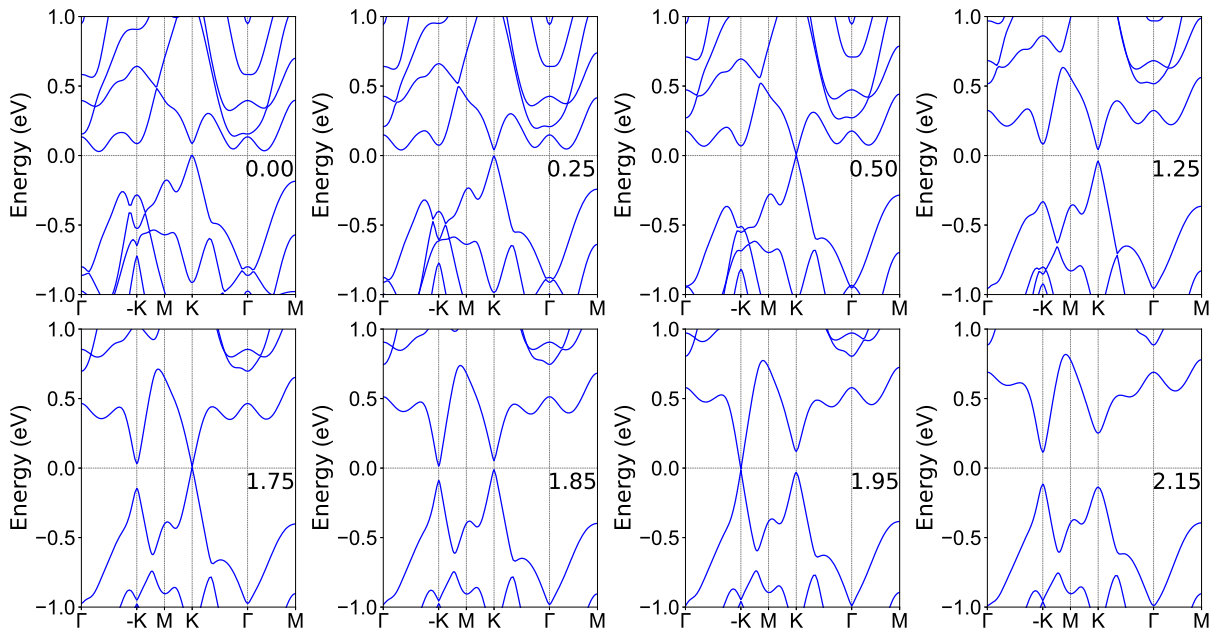}
  \caption{(Color online) The energy band structures of $\mathrm{OsBr_2}$ monolayer with out-of-plane  magnetic anisotropy  by using GGA+SOC at representative $U$ values. }\label{u-band}
\end{figure*}

\begin{figure}
   \includegraphics[width=8cm]{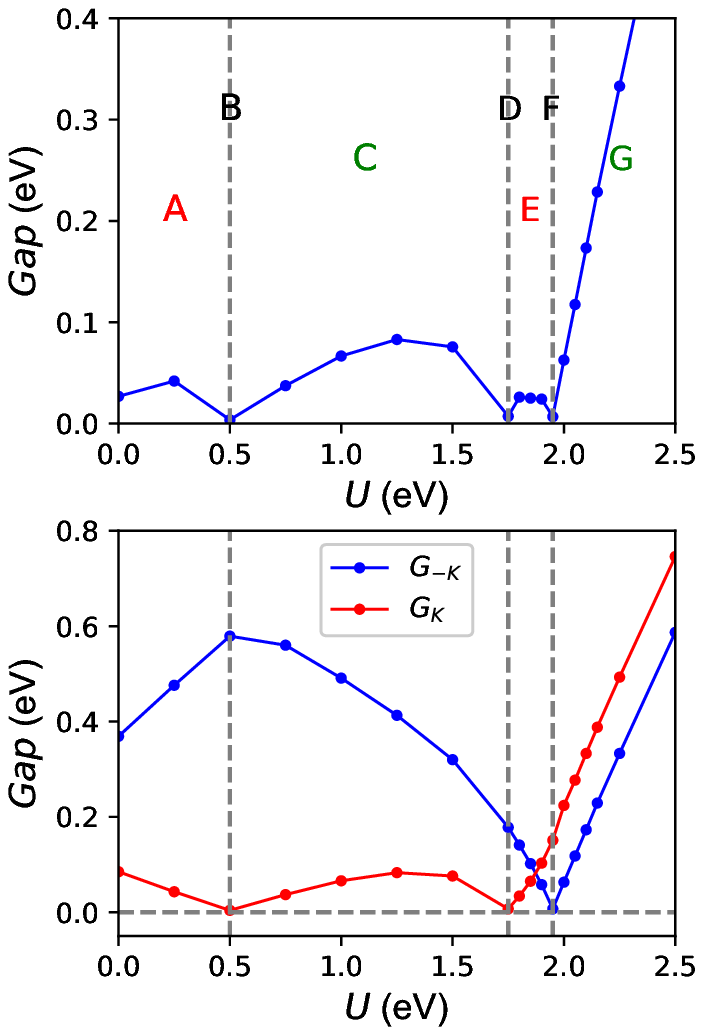}
  \caption{(Color online)For $\mathrm{OsBr_2}$ monolayer with out-of-plane  magnetic anisotropy, the  global energy band gap and  energy  band gaps for -K and K valleys as a function of   $U$. A and E regions mean PVQAHI, and C and G regions mean PFVI, and B, D and F points mean PHVM. }\label{u-gap}
\end{figure}

\section{Computational detail}
The spin-polarized  first-principles calculations  are carried out  within density-functional
theory  (DFT)\cite{1},  as implemented in VASP code\cite{pv1,pv2,pv3}. The projected
augmented wave (PAW) method with  GGA of Perdew-Burke-Ernzerhof (PBE-GGA)\cite{pbe} exchange-correlation functional  is adopted.
The energy cut-off of 500 eV  and total energy  convergence criterion of  $10^{-8}$ eV are used to attain accurate results. The force
convergence criteria on each atom  is set to be less than 0.0001 $\mathrm{eV.{\AA}^{-1}}$.
A more than 18 $\mathrm{{\AA}}$ cell height is used in
the $z$ direction to prevent periodic images from interacting with
each other.
 We use $\Gamma$-centered 24$\times$24$\times$1 k-mesh to sample the Brillouin zone (BZ) for structure optimization, electronic structures and elastic stiffness tensor, and 12$\times$24$\times$1 Monkhorst-Pack k-point mesh for FM/antiferromagnetic (AFM)  energy and piezoelectric stress tensor  with rectangle supercell.
Within  the rotationally invariant approach proposed by Dudarev et al, the   GGA+$U$  method  is employed to describe the correlated  Os-$d$ electrons. The SOC effect is explicitly included, which is very key to investigate magnetic anisotropy energy (MAE), electronic and topological properties of $\mathrm{OsBr_2}$ monolayer.

We use strain-stress relationship (SSR) and density functional perturbation theory (DFPT) method\cite{pv6} to attain  elastic stiffness tensor  $C_{ij}$ and piezoelectric stress tensor $e_{ij}$.
The  2D elastic/piezoelectric  coefficients $C^{2D}_{ij}$/$e^{2D}_{ij}$ have been renormalized by   $C^{2D}_{ij}$/$e^{2D}_{ij}$=$L_z$$C^{3D}_{ij}$/$e^{3D}_{ij}$, where  the $L_z$  is  the cell height along $z$ direction. The  edge states  are calculated with the maximal localized
Wannier function tight-binding model by employing $d$-orbitals of Os atoms and  $p$-orbitals of Br atoms\cite{w1,w2}.
The Berry curvatures of $\mathrm{OsBr_2}$
are attained directly from the calculated
wave functions  based on Fukui's
method\cite{bm}, as implemented in VASPBERRY code\cite{bm1,bm2}. For predicting Curie temperature ($T_C$) of  $\mathrm{OsBr_2}$, the 40$\times$40 supercell and  $10^7$ loops are used to achieve
Monte Carlo (MC) simulations, as implemented in Mcsolver code\cite{mc}.

\section{Structure and magnetic properties}
Similar to monolayer 1$H$-$\mathrm{MoS_2}$, for monolayer $\mathrm{OsBr_2}$, its Os atom layer is sandwiched by two
Br atom layers through the Os-Br bonds, whose crystal structure is shown in \autoref{st}, along with BZ with high-symmetry points in FIG.1 of electronic supplementary information (ESI).
The $\mathrm{OsBr_2}$ monolayer with $P\bar{6}m2$  symmetry (No.187) lacks centrosymmetry, indicating that it should  possess piezoelectricity and FV properties.  The symmetry of $\mathrm{OsBr_2}$ is higher than that of  FeClF with  $P3m1$  symmetry (No.156) due to broken vertical mirror symmetry\cite{q12}.
These mean that only in-plane piezoelectric polarization exists in $\mathrm{OsBr_2}$ monolayer,  when it is subject to a uniaxial in-plane strain.   When applying biaxial in-plane strain, the in-plane piezoelectric polarization will be suppressed.
The lattice constants $a$ of $\mathrm{OsBr_2}$ monolayer is optimized with varied $U$, as shown in FIG.1 of ESI. The $a$ increases with increasing $U$, which can also be found in  monolayer FeClF and  $\mathrm{RuBr_2}$\cite{q12,q13}. And then, its magnetic, electronic and piezoelectric properties are investigated with varied $U$ by using the corresponding $a$.

Next, a rectangle supercell (see \autoref{st}) is used to explore the magnetic coupling of $\mathrm{OsBr_2}$ monolayer, and
two initial magnetic configurations of  AFM and FM ordering are considered. The AFM
configuration is  plotted in FIG.1 of ESI, and similar AFM
configuration is used in  monolayer $\mathrm{VSi_2P_4}$\cite{q11}. Their
total energy difference $\Delta E$ between AFM and FM ordering as a function of $U$ is plotted in \autoref{u-em}. In considered $U$ range, $\mathrm{OsBr_2}$ monolayer is a 2D
intrinsic FM material. It is found that $\Delta E$ is sensitive to $U$, giving rise to important influence on $T_C$ of  $\mathrm{OsBr_2}$ monolayer.
In view of the important role of magnetic anisotropy in realizing the long-range
magnetic order and novel electronic states in 2D materials\cite{q11,q12,q13}, the MAE of $\mathrm{OsBr_2}$ monolayer is calculated from a difference in the obtained total energies with magnetization
direction parallel or perpendicular to the plane of monolayer ($E_{MAE}$ = $E_{(100)}$-$E_{(001)}$).  Thus, the  positive or negative  MAE means
that the easy magnetization axis is perpendicular or parallel to the plane of monolayer. The MAE vs $U$ is plotted in \autoref{u-em}, and the easy magnetization axis changes from out-of-plane to in-plane one with critical $U$ value about 2.6 eV. Similar results can be found in monolayer FeClF and $\mathrm{RuBr_2}$\cite{q12,q13}, but critical $U$ value of $\mathrm{OsBr_2}$ is larger than their ones (1.45 and 2.07 eV). The large critical $U$ value is very important to confirm intrinsic novel QAH and HVM states.
 It is well known that  valence 5$d$ wave functions are more delocalized
than those of 3$d$, and then 5$d$ transition-metals show very weak electron correlation.  In previous works, the  $U$ for  Os-5$d$ electrons
is taken as  0.5 eV for monolayer $\mathrm{OsO_2}$ and   1.5 eV for the 5$d$-modified  antimonene\cite{q15,q16}. At  typical $U=$1.5 eV,  the MAE
of monolayer $\mathrm{OsBr_2}$  is as high as 16.14 meV/Os, which means very stable  out-of-plane  magnetic anisotropy.

\begin{figure}
  \includegraphics[width=8cm]{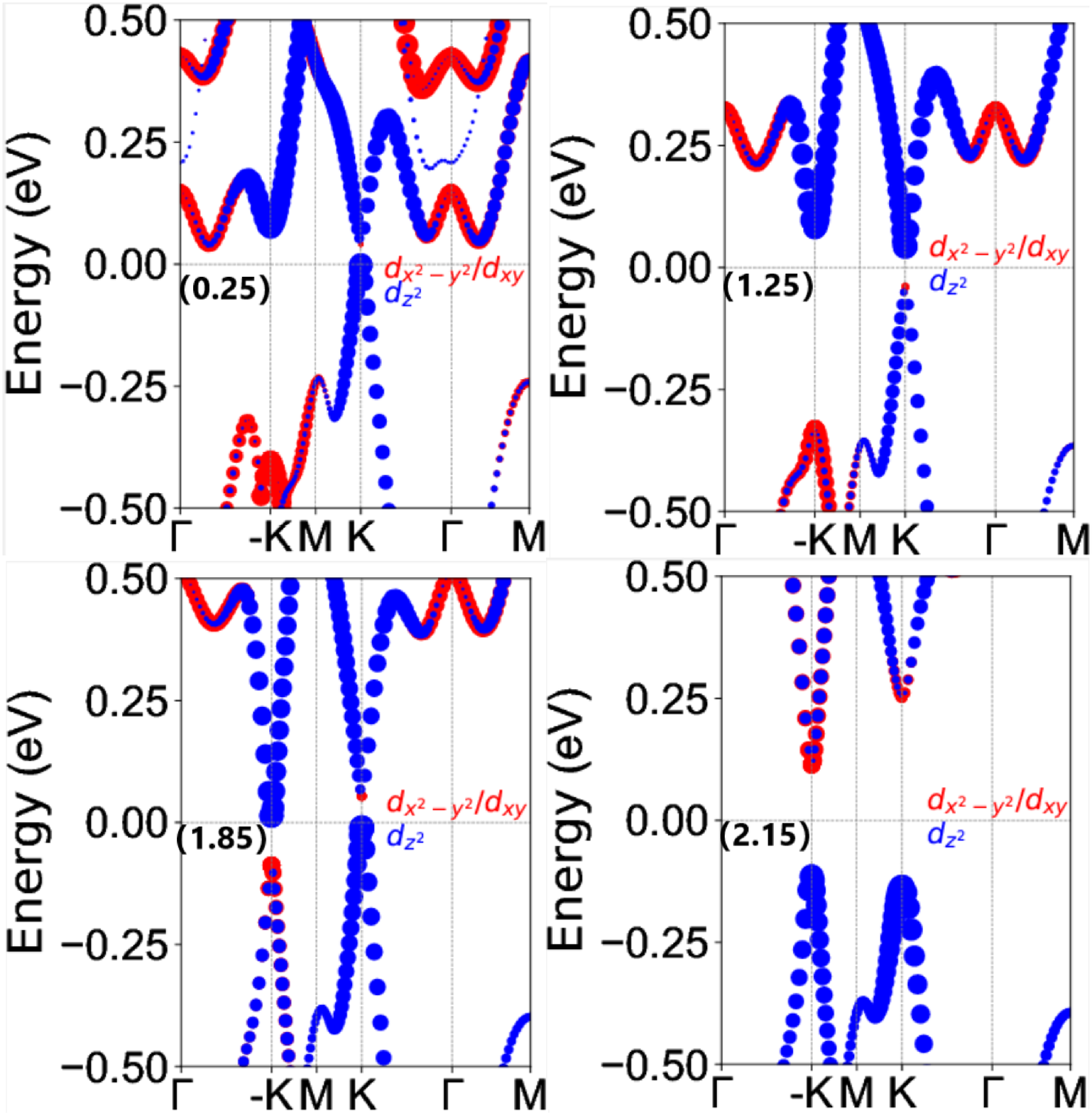}
\caption{(Color online)For $\mathrm{OsBr_2}$ monolayer with out-of-plane  magnetic anisotropy,, the Os-$d_{x^2-y^2}$/$d_{xy}$ and $d_{z^2}$-orbital characters energy band structures at representative  $U$$=$0.25 eV, 1.25 eV, 1.85 eV and 2.15 eV.}\label{u-p}
\end{figure}

\begin{figure*}
  \includegraphics[width=16cm]{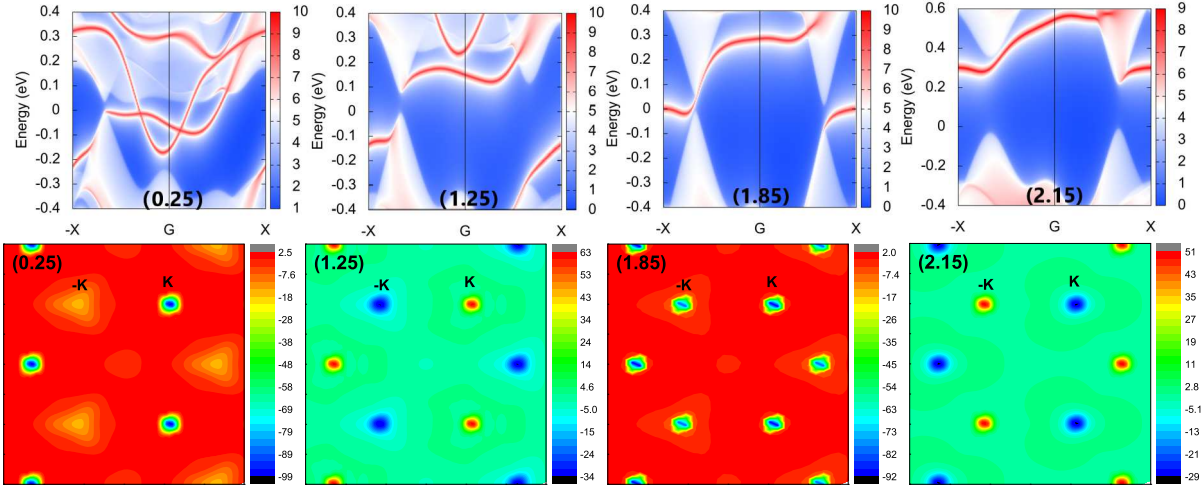}\\
\caption{(Color online)For $\mathrm{OsBr_2}$ monolayer with out-of-plane  magnetic anisotropy,  the topological
edge states (Top) and Berry curvature distribution in 2D BZ (Bottom) at representative  $U$$=$0.25 eV, 1.25 eV, 1.85 eV and 2.15 eV.}\label{u-sb}
\end{figure*}

\begin{figure}
  \includegraphics[width=7cm]{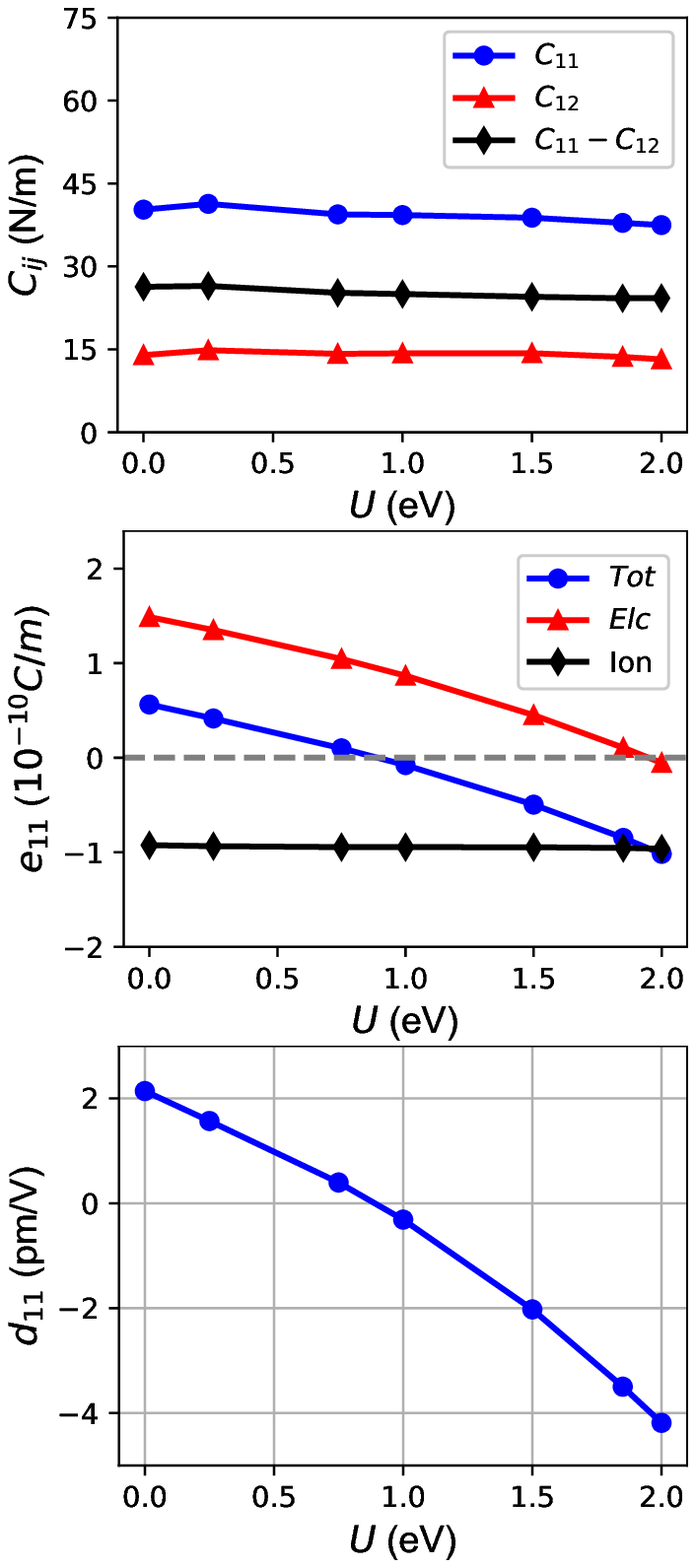}
  \caption{(Color online)For $\mathrm{RuBr_2}$ monolayer,  the elastic constants $C_{ij}$, piezoelectric stress coefficient $e_{11}$ along with ionic and electronic parts, and  piezoelectric strain coefficient $d_{11}$  as a function of $U$.}\label{u-ed}
\end{figure}

\section{topological phase transition}
Electronic correlation combined with  out-of-plane magnetic anisotropy can produce novel electronic states in some 2D materials\cite{q10,q11,q12,q13}, such as FV, QAH and HVM states. However, the in-plane magnetic anisotropy will lead to disappeared novel electronic states in these 2D materials\cite{q11,q12,q13}.
Generally, Os-5$d$ electrons show very weak electron correlation\cite{q15,q16}, and the easy magnetization axis of  $\mathrm{OsBr_2}$ monolayer is out-of-plane with $U$ being less than  2.6 eV. So, we only consider that the $U$ ranges from 0.00 eV to 2.50 eV, and the corresponding electronic properties are investigated.  At some representative $U$ values,  the energy band structures of  $\mathrm{OsBr_2}$ with GGA+SOC  are shown in \autoref{u-band}, and
the evolutions of total energy band gap along with those at -K/K point  as a function of $U$ are plotted in \autoref{u-gap}.

It is clearly seen that there are three points around $U$$=$0.5 eV, 1.75 eV and 1.95 eV, where the total energy band gap is closed.
At these points,  the HVM  state can be achieved, whose  conduction electrons are intrinsically 100\% valley polarized\cite{q10}.
However, these HVM states can be divided into two categories. At $U$$=$0.5 eV/1.75 eV,  the band gap gets closed  at K valley,
 while a  band gap of 0.58 eV/0.18 eV  is kept at -K valley. At $U$$=$1.95 eV, the band gap of -K valley is closed, while the band gap at K valley is 0.15 eV.  The $U$ region can be divided into four parts by three HVM states.

  As shown in \autoref{u-p}, both  valence and conduction
valleys at -K and K points are primarily contributed by Os-$d_{x^2-y^2}$/$d_{xy}$ or $d_{z^2}$  orbitals in considered $U$ range.
For 0.00 eV$<$$U$$<$0.50 eV,  the $d_{x^2-y^2}$ and $d_{xy}$ orbitals dominate   conduction band at K valley, while the valence band of K valley is mainly from $d_{z^2}$ orbitals (For example $U$$=$0.25 eV).  When $U$ is between 0.5 eV and 1.75 eV, the opposite situation can be observed with ones of 0.00 eV$<$$U$$<$0.50 eV (For example $U$$=$1.25 eV). For 1.75 eV$<$$U$$<$1.95 eV, the distribution of  $d_{x^2-y^2}$/$d_{xy}$ and  $d_{z^2}$ orbitals at K valley is opposite to one of 0.50 eV$<$$U$$<$1.75 eV (For example $U$$=$1.85 eV). For the three regions, at -K valley,  the $d_{z^2}$  orbitals dominate   conduction band, while the valence band is mainly from  $d_{x^2-y^2}$/$d_{xy}$ orbitals. These means that there are two-time band inversion  between $d_{xy}$/$d_{x^2-y^2}$ and $d_{z^2}$ orbitals at K valley  with increasing $U$.  For the fourth region ($U$$>$1.95 eV), at -K valley, the distribution of  $d_{x^2-y^2}$/$d_{xy}$ and  $d_{z^2}$ orbitals is opposite to one of 1.75 eV$<$$U$$<$1.95 eV, but this is the same at K valley (For example $U$$=$2.15 eV). This means another band inversion  between $d_{xy}$/$d_{x^2-y^2}$ and $d_{z^2}$ orbitals at -K valley.

The three HVM states imply  that the total gap of $\mathrm{OsBr_2}$ monolayer closes  and reopens  three times. The  special gap change along with band inversion suggest some topological phase transitions, and  QAH state  may exist in some regions.
To confirm  QAH  phases,  we calculate the  edge states at representative  $U=$0.25 eV, 1.25 eV, 1.85 eV and 2.15 eV from four regions, which are plotted in \autoref{u-sb}. It is clearly seen that a nontrivial chiral edge state, connecting  the conduction bands and
valence  bands,  does exist in two regions (0.00 eV$<$$U$$<$0.50 eV and 1.75 eV$<$$U$$<$1.95 eV), implying a QAHI.
The calculated  Chern number is equal to minus one ($C$=-1), which is consistent with one obtained
by integrating the Berry curvature within the first BZ.
For the other two regions (0.50 eV$<$$U$$<$1.75 eV and $U$$>$1.95 eV), no nontrivial chiral edge state appears, suggesting a normal FM semiconductor.
These mean that increasing $U$ can induce threefold  topological phase transition in monolayer $\mathrm{OsBr_2}$

These  topological phase transitions are also connected with transformation of Berry curvature.
At representative  $U=$0.25 eV, 1.25 eV, 1.85 eV and 2.15 eV, the distributions of Berry curvature are shown in \autoref{u-sb}, whose hot spots  are around -K and K valleys. For two regions (0.50 eV$<$$U$$<$1.75 eV and $U$$>$1.95 eV), the opposite
signs and different magnitudes around -K and K valleys can be observed. However, for the other two regions (0.00 eV$<$$U$$<$0.50 eV and 1.75 eV$<$$U$$<$1.95 eV), the
 Berry curvatures  around -K and K valleys have the same signs and different magnitudes.
 With increasing $U$, triple topological phase transitions are produced, which are connected by three HVM states.  In these transitions, the sign of Berry curvature  at -K or K valley will  flip. For example the first two topological phase transitions,    the negative
  Berry curvature of K valley ($U$$=$0.25 eV)  changes into positive one  ($U$$=$1.25 eV), and then changes into negative one ($U$$=$1.85 eV).
The third  topological phase transition  leads to  the sign flipping  of Berry curvature  at -K valley, and the negative  Berry curvature ($U$$=$1.85 eV) changes into positive one  ($U$$=$2.15 eV). These suggest  that sign-reversible  Berry curvature can be induced by  electronic correlation, which is related with  topological phase transition.

\section{Piezoelectric properties}
Similar to  monolayer $\mathrm{MoS_2}$\cite{p1},  monolayer $\mathrm{OsBr_2}$ lacks  inversion symmetry, but possesses a reflection symmetry with respect to the central Os atomic plane. This means only
 in-plane polarization along the armchair direction  is allowed when $\mathrm{OsBr_2}$ is subject to a uniaxial in-plane strain. For biaxial in-plane strain, the in-plane piezoelectric polarization will be suppressed. The third-rank piezoelectric stress tensor  $e_{ijk}$ and strain tensor $d_{ijk}$ are defined as:
 \begin{equation}\label{pe0}
      e_{ijk}=\frac{\partial P_i}{\partial \varepsilon_{jk}}=e_{ijk}^{elc}+e_{ijk}^{ion}
 \end{equation}
and
 \begin{equation}\label{pe0-1}
   d_{ijk}=\frac{\partial P_i}{\partial \sigma_{jk}}=d_{ijk}^{elc}+d_{ijk}^{ion}
 \end{equation}
in which $P_i$, $\varepsilon_{jk}$ and $\sigma_{jk}$ are polarization vector, strain and stress, respectively.
The $e_{ijk}^{elc}$/$d_{ijk}^{elc}$ means  clamped-ion piezoelectric coefficients with only  considering electronic contributions.  The  $e_{ijk}$/$d_{ijk}$ means relax-ion piezoelectric coefficients as a realistic result, which is from the sum of ionic ($e_{ijk}^{ion}$/$d_{ijk}^{ion}$)
and electronic ($e_{ijk}^{elc}$/$d_{ijk}^{elc}$) contributions.
Analogous to monolayer $\mathrm{MoS_2}$\cite{p1}, the  piezoelectric stress   and strain tensors of $\mathrm{OsBr_2}$ by using  Voigt notation  can be reduced into:
  \begin{equation}\label{pe1}
  \left(
    \begin{array}{ccc}
      e_{11} &-e_{11} & 0 \\
    0 &0 & -e_{11}\\
      0 & 0 & 0 \\
    \end{array}
  \right)
  \end{equation}
  \begin{equation}\label{pe1}
  \left(
    \begin{array}{ccc}
        d_{11} & -d_{11} & 0 \\
    0 &0 & -2d_{11} \\
      0 & 0 & 0 \\
    \end{array}
  \right)
  \end{equation}
The only independent $d_{11}$  can be attained  by $e_{ik}=d_{ij}C_{jk}$:
\begin{equation}\label{pe2}
    d_{11}=\frac{e_{11}}{C_{11}-C_{12}}
\end{equation}

Firstly, we calculate elastic stiffness tensor  $C_{ij}$ of $\mathrm{OsBr_2}$ at some representative $U$ values, which are plotted in \autoref{u-ed}.
For $\mathrm{OsBr_2}$ monolayer, the  Born  criteria of mechanical stability\cite{ela} ($C_{11}$$>$0 and $C_{11}-C_{12}$$>$0) is satisfied for all $U$ values,    indicating  its  mechanical stability. It is found that $C_{ij}$ have weak dependence on $U$.
The  $e_{11}$  of monolayer $\mathrm{OsBr_2}$ is calculated with orthorhombic supercell by using DFPT method at some representative $U$ values.
 The piezoelectric  stress  coefficients  $e_{11}$  (including ionic  and electronic contributions)  and piezoelectric  strain  coefficients $d_{11}$  are plotted in \autoref{u-ed}. When $U$ is less than about 1.85 eV, the electronic and ionic polarizations  have  opposite signs.
 For $U$$<$1.0 eV, the electronic contribution dominates the in-plane piezoelectricity. For 1.0 eV$<$$U$$<$1.85 eV, the ionic part dominates the $e_{11}$. According to \autoref{pe2}, we calculate $d_{11}$ from previous calculated $C_{ij}$ and $e_{11}$.
 With increasing $U$,  the $d_{11}$ changes from positive value to negative one, and the trend is the same with $e_{11}$.
At representative $U$$=$1.5 eV, the absolute value of $d_{11}$ is 2.02 pm/V, which is  close to one of  $\alpha$-quartz ($d_{11}$=2.3 pm/V).

\section{electronic states}
\begin{figure}
  \includegraphics[width=8cm]{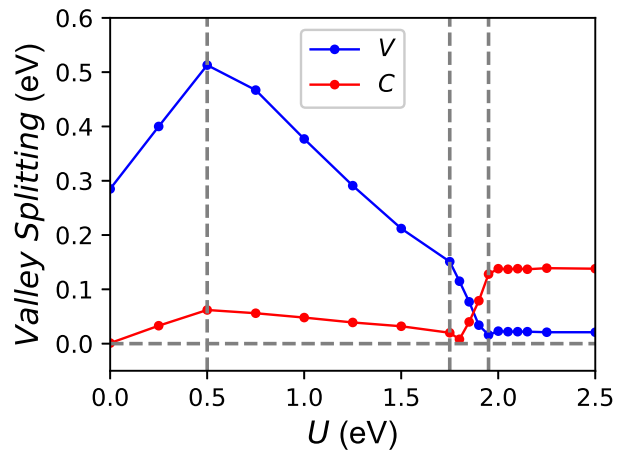}
\caption{(Color online)For  out-of-plane magnetic anisotropy, the absolute value of valley splitting of monolayer $\mathrm{OsBr_2}$  in both conduction ($C$) and valence ($V$) bands as a function of $U$.}\label{v-v}
\end{figure}

In considered $U$ range, $\mathrm{OsBr_2}$  is a FV material, and the valley splitting for both valence and condition bands is plotted in \autoref{v-v}.
 A possible way has been proposed  to realize anomalous valley Hall effect  in monolayer $\mathrm{GdCl_2}$ by piezoelectric effect\cite{gsd1}, not an  external
electric field.  The $\mathrm{OsBr_2}$  monolayer has the same structure with $\mathrm{GdCl_2}$, and has FV and piezoelectric properties. So,  $\mathrm{OsBr_2}$  is also a piezoelectric FV  (PFV) material, which can be used to realize piezoelectric anomalous valley Hall effect (PAVHE), as is
illustrated in FIG.2 of ESI.  The in-plane longitudinal electric field $E$ is induced  with an  applied   uniaxial in-plane strain by piezoelectric effect, and then anomalous valley Hall effect can be produced,
The $U$  should be determined from future experiment result. If the $U$ falls  into the two regions (0.00 eV$<$$U$$<$0.50 eV and 1.75 eV$<$$U$$<$1.95 eV), $\mathrm{OsBr_2}$  monolayer will possess FV, QAH and piezoelectric properties, namely PVQAHI. For 0.00 eV$<$$U$$<$0.50 eV, chiral gapless edge mode mixes with trivial edge state in bulk gap, but the pure nontrivial chiral edge state can be observed for 1.75 eV$<$$U$$<$1.95 eV.
In case of $U$=0.50 eV, 1.75 eV or 1.95 eV,  $\mathrm{OsBr_2}$  has HVM and piezoelectric properties, namely PHVM.  These provide possibility to tune QAH and anomalous valley Hall effects by piezoelectric effect.

\begin{figure*}
  \includegraphics[width=15cm]{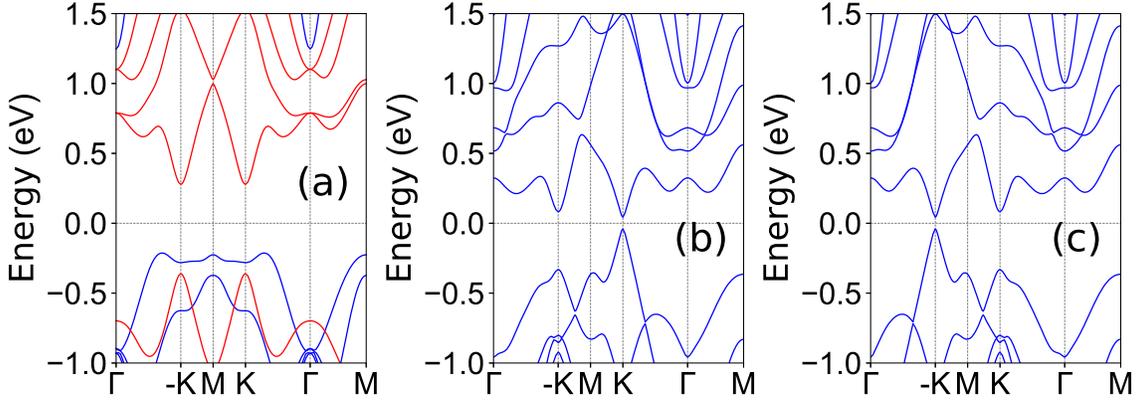}
\caption{(Color online)For out-of-plane magnetic anisotropy, the band structure of  monolayer  $\mathrm{OsBr_2}$ (a) without SOC; (b) and (c) with SOC for magnetic moment of Os along the positive and negative $z$ direction, respectively. In (a), the blue (red) lines represent the band structure in the spin-up (spin-down) direction.}\label{band-z}
\end{figure*}

\begin{figure*}
  \includegraphics[width=12cm]{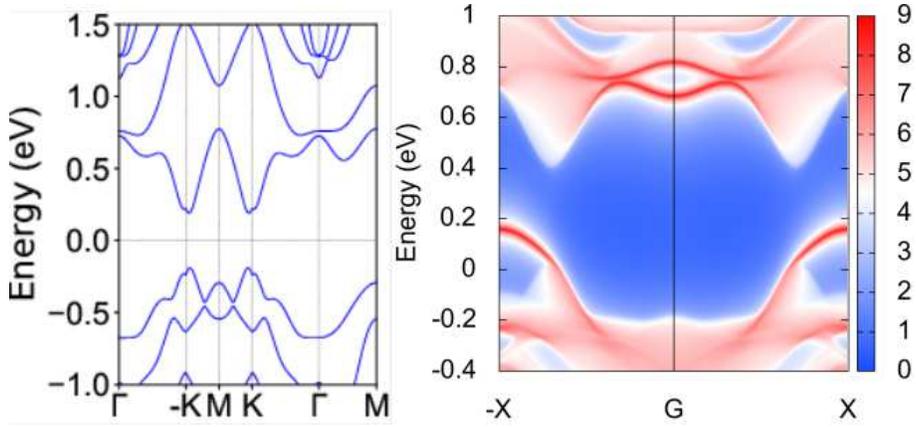}\\
\caption{(Color online)For $\mathrm{OsBr_2}$ monolayer with in-plane  magnetic anisotropy,  the energy band structures (Left) and topological
edge states (Right)  at representative  $U$$=$1.85 eV.}\label{u-sb1}
\end{figure*}

When reversing the magnetization orientation, the valley polarized state is also reversed. To explain this,  the spin-polarized energy band structures of monolayer  $\mathrm{OsBr_2}$  are shown in \autoref{band-z} without SOC and with SOC for magnetic moment of Os along the positive and negative $z$ direction at representative $U$=1.25 eV. Without SOC, the bottom conduction band is from
the spin-down channel,  and there are a pair
of energy extremes  at  -K and K points, yielding
two inequivalent but degenerate valleys. However,
 the top valance band comes from
the spin-up channel, and no energy extremes appear at  -K and K points. In fact, these results depend on $U$ value.
Increasing $U$ can lead to  that both bottom conduction and top valance bands are from
the spin-down channel, and have a pair
of energy extremes  at  -K and K points, which are inequivalent but degenerate valleys (see FIG.3 of ESI at $U$$=$2.15 eV). When considering SOC, there are a pair
of energy extremes  at  -K and K points for both conduction and valence bands, and
the valley degeneracy is lifted (the -K/K valley state has a
lower energy than K/-K valley for valence/conduction bands.), producing  valley polarized
state in  $\mathrm{OsBr_2}$. It is found that the valley polarization
of valence bands is remarkably larger than one of conduction bands, which is because  the -K and K valleys of valence bands are dominated by $d_{x^2-y^2}$ and $d_{xy}$ orbitals, while those of conduction bands are mainly from  $d_{z^2}$ orbitals. Similar phenomenons can be found in many 2D FV materials\cite{q11,q12,q13,v2,v3}.   As shown in \autoref{band-z} (c), the  valley polarization
can be flipped  by reversing the magnetization
of Os atoms,  namely,  the K (-K) valley state has a
lower energy than -K (K) valley for valence (conduction) bands. Manipulating  direction of
magnetization of  $\mathrm{OsBr_2}$ may be  an efficient way to tune its valley
properties.

The different magnetic orientation will affect the symmetry of $\mathrm{OsBr_2}$, which has important influence on
its electronic properties. For in-plane  magnetic anisotropy, the FV and QAH properties will disappear. At representative  $U=$1.85 eV,
with in-plane  magnetic anisotropy,  the energy band structures  and topological
edge states by using GGA+SOC are plotted in \autoref{u-sb1}. It is clearly seen that the energies at -K and K points are degenerate for both valence and conduction bands,  giving rise to no valley polarized
state.  The edge-state calculations show no chiral gapless edge modes within the bulk gap. So,  the intrinsic out-of-plane magnetic anisotropy is very important to confirm these novel electronic states and topological transformations in considered $U$ range (0.00-2.50 eV).

\section{Curie temperature}
 As shown in \autoref{u-em}, the electronic correlation effects ($U$) have important influence on  the strength of  FM interaction, which is related with   $T_C$ of monolayer $\mathrm{OsBr_2}$. The $T_C$ is estimated  based on the Heisenberg model by
MC simulations within Wolf algorithm. An effective classical spin Heisenberg
model can be written as:
  \begin{equation}\label{pe0-1-1}
H=-J\sum_{i,j}S_i\cdot S_j-A\sum_i(S_i^z)^2
 \end{equation}
where  $S_i$/$S_j$, $S_i^z$,  $J$ and  $A$  are   the
spin vectors of each Os atom, spin component parallel to the $z$ direction,  the nearest neighbor exchange parameter and   MAE, respectively.
With rectangle supercell, the total energies of  AFM  ($E_{AFM}$) and FM ($E_{FM}$) ordering  with normalized spin vector ($|S|$=1) are given
as:
 \begin{equation}\label{pe0-1-2}
E_{FM}=E_0-6J-2A
 \end{equation}
  \begin{equation}\label{pe0-1-3}
E_{AFM}=E_0+2J-2A
 \end{equation}
where $E_0$ is the total energy of systems without magnetic coupling.
The  $J$ can be written as:
  \begin{equation}\label{pe0-1-3}
J=\frac{E_{AFM}-E_{FM}}{8}
 \end{equation}

\begin{figure}
  \includegraphics[width=8cm]{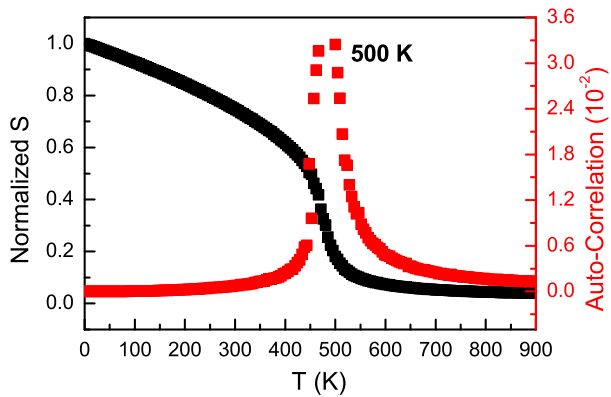}
  \caption{(Color online)For  monolayer  $\mathrm{OsBr_2}$, the normalized magnetic moment (S) and auto-correlation  as a function of temperature with $U$$=$2.5 eV.}\label{tc}
\end{figure}

   According to \autoref{u-em},   increasing $U$ weakens FM interaction, which will reduce $T_C$. We take $U$$=$2.50 eV as a representative case, and
   the calculated normalized $J$ is 35.54 meV.
 The   normalized magnetic moment and auto-correlation  as a function of temperature  are plotted in \autoref{tc}, and the predicted $T_C$ is about 500 K. In considered $U$ range (0.00-2.50 eV), the predicted $T_C$ is all higher than room temperature. This implies that
$\mathrm{OsBr_2}$  is indeed a room-temperature ferromagnet.

\section{Discussion and Conclusion}
\begin{figure}
  \includegraphics[width=8cm]{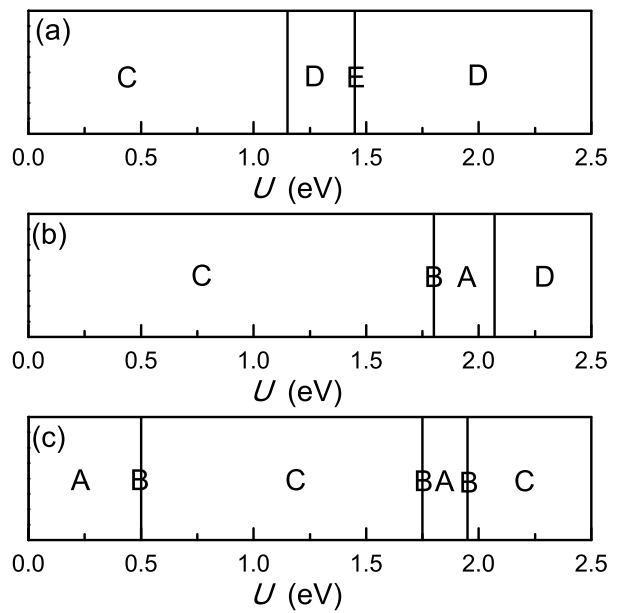}
  \caption{The intrinsic phase diagrams for monolayer FeClF (a), $\mathrm{RuBr_2}$ (b) and $\mathrm{OsBr_2}$ (c) with different $U$ values: A region means QAHI, B
  point  means HVM, C region means FVI, D region means common FM semiconductor, and E point means common FM semimetal.}\label{ip}
\end{figure}

Although Fe, Ru and Os atoms have same outer valence electrons, their SOC strengths are different due to different atomic mass.
Increasing  $U$ along with different SOC strength can induce different phase diagram of electronic state with fixed out-of-plane  magnetic anisotropy.
For monolayer $\mathrm{FeCl_2}$ and $\mathrm{RuBr_2}$, twofold  topological phase transition with fixed out-of-plane  magnetic anisotropy can be induced with increasing $U$\cite{q10,q13}, and  the order is FVI to HVM to  VQAHI to HVM to FVI. However, for  $\mathrm{OsBr_2}$, threefold  topological phase transition can be observed, and it  undergoes   VQAHI, HVM,  FVI, HVM, VQAHI, HVM,  FVI, when $U$ increases. Strong SOC can lead to high critical $U$ value of out-of-plane to in-plane transition, which is very important to produce novel phase diagram. For example, for FeClF monolayer, the intrinsic  phase diagram shows no special QAH and HVM states due to small critical $U$  (about 1.15 eV)\cite{q12}. However, intrinsic  phase diagram of  $\mathrm{OsBr_2}$ shows both special QAH and HVM states with large critical $U$ of about 2.6 eV. The intrinsic phase diagrams for monolayer FeClF, $\mathrm{RuBr_2}$ and $\mathrm{OsBr_2}$ are plotted in \autoref{ip}. It is clearly seen that the intrinsic phase diagram of $\mathrm{OsBr_2}$ is different from those of monolayer FeClF and $\mathrm{RuBr_2}$.

The importance of electron correlations has been proved  on the electronic state  of monolayer $\mathrm{OsBr_2}$.
The different  correlation strength  (varied $U$) can give rise to different electronic state. For a given material, the correlation strength should be fixed, and $\mathrm{OsBr_2}$ should belong to a particular  electronic state in the
phase diagram, which should  be determined from related experiment. However,  varied $U$ in producing novel electronic state is equivalent to applying different strain, which has been confirmed in $\mathrm{RuBr_2}$ monolayer\cite{q13}. With fixed out-of-plane  magnetic anisotropy for $\mathrm{RuBr_2}$, the phase diagram with different $U$ values is similar with one with different strain. So, the rich electronic state and novel phase transitions can still be achieved in practice by strain. In fact, dual topological phase transition has been achieved in monolayer $\mathrm{OsBr_2}$ by strain\cite{q14}.
The sign-reversible valley-dependent Berry phase effects and QAH/HVM states in septuple atomic  monolayer $\mathrm{VSi_2N_4}$  has been achieved by strain\cite{v6}.

In summary, we have demonstrated threefold  topological phase transition with different $U$  in monolayer $\mathrm{OsBr_2}$, which are related with sign-reversible  Berry curvature and band inversions of $d_{xy}$/$d_{x^2-y^2}$ and $d_{z^2}$ orbitals at -K and K valleys.  In considered $U$ range (0.00-2.50 eV), $\mathrm{OsBr_2}$ is an intrinsic FVI with constant out-of-plane  magnetic anisotropy.
 There are two  QAH phase regions characterized
by a  chiral gapless edge mode, and the second region has pure non-trivial edge mode without mixture of characterless edge mode.
The right  boundary of the first QAH phase region and  two boundaries  of the second QAH phase region   correspond to the HVM with fully valley
polarized carriers. Due to lacking  inversion symmetry, $\mathrm{OsBr_2}$ is piezoelectric with only independent $d_{11}$, which provides  possibility
to achieve anomalous valley Hall effect  by piezoelectric effect. The estimated high $T_C$ confirm that these possible novel states can be realized
in the high temperature. Our works deepens our understanding of  strong SOC
 combined with  correlation effects in monolayer $\mathrm{OsBr_2}$, and provide a platform for multifunctional 2D material, such as PVQAHI and PFVI.

\begin{acknowledgments}
This work is supported by Natural Science Basis Research Plan in Shaanxi Province of China  (2021JM-456),  Graduate Innovation Fund Project in Xi'an University of Posts and Telecommunications (CXJJDL2021001),  the Nature Science Foundation of China (Grant No.11974393) and the Strategic Priority Research Program of the Chinese Academy of Sciences (Grant No. XDB33020100). We are grateful to the Advanced Analysis and Computation Center of China University of Mining and Technology (CUMT) for the award of CPU hours and WIEN2k/VASP software to accomplish this work.
\end{acknowledgments}

\end{document}